\documentclass[aps,twocolumn]{revtex4-1}
\usepackage{amsmath,amssymb}
\usepackage{slashed}
\usepackage{graphicx}
\usepackage{bm}
\usepackage[T1]{fontenc}
\usepackage[utf8]{inputenc}
\usepackage{gauss} 
\usepackage[top=1.0in,bottom=1.0in,left=1.0in,right=1.0in]{geometry}
\usepackage[colorlinks,linkcolor=blue,citecolor=blue]{hyperref}
\usepackage{comment}
\newcommand{\be}{\begin{equation}}
\newcommand{\ee}{\end{equation}}
\newcommand{\bea}{\begin{eqnarray}}
\newcommand{\eea}{\end{eqnarray}}
\newcommand{\ba}{\begin{eqnarray}}
\newcommand{\ea}{\end{eqnarray}}

\begin{document}

\title{Four-nucleon clustering near the QCD critical point: \\
theory versus experiment
}

\author{Edward Shuryak}
\email{edward.shuryak@stonybrook.edu}
\affiliation{Center for Nuclear Theory, Department of Physics and Astronomy, Stony Brook University, Stony Brook, New York 11794--3800, USA}

\begin{abstract} 
This letter relates theoretical predictions made
previously with recently announced RHIC BES-II
experimental data on multi-baryon fluctuations.
We conclude that new 
statistically significant effect is observed, 
and its features are in good agreement with predictions.
 \end{abstract}

\maketitle
\section{Theory of four baryon clusters}

The original suggestion \cite{Stephanov:1998dy} to search for the QCD critical point (CP) via energy scan 
with focus on event-by-event fluctuations resulted in 
BES program at RHIC. Out of large theoretical 
literature we focus here on  a sequence of papers focusing
on ``kurtosis" of nucleon multiplcity distribution
and its relation to four-nucleon correlations at freezeout conditions. 

In \cite{Shuryak:2018lgd,Shuryak:2019ikv,DeMartini:2020anq} we used classical molecular dynamics, novel semiclassical ``flucton" method,
and then direct Path Integral Monte Carlo, with the usual nuclear forces as well as their versions modified near CP. All of those focus
on four nucleon systems, the lightest one which possesses ``preclusters" decaying into multiple ($\sim 50$) bound states with small binding, as well as small positive energy states. To our
knowledge no other group related event-by-event and thermodynamical
susceptibilities to local clustering of the nucleons.

\begin{figure}[h!]
        \centering
        \includegraphics[width=0.6\linewidth]{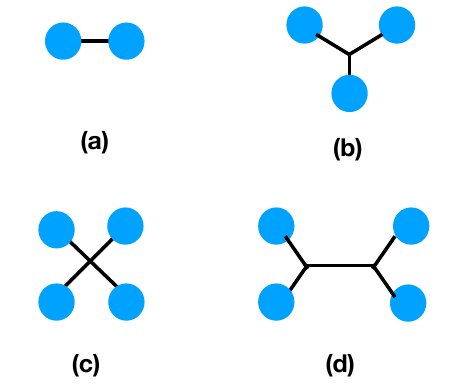}
        \caption{Two,three and four-body interactions of the nucleons due to exchanges of the ``order parameter" field $\phi$}
        \label{fig_4diagrams}
    \end{figure}

The modification near CP is due to novel interactions between
nucleons caused by the exchanges of the ``order parameter field" with large correlation length
near the CP. This field has nonlinear coupling with itself,
producing three and four-body interactions between nucleons  shown in Fig.\ref{fig_4diagrams}.  

\begin{figure}[h!]    
        \centering
    \includegraphics[width=0.7\linewidth]{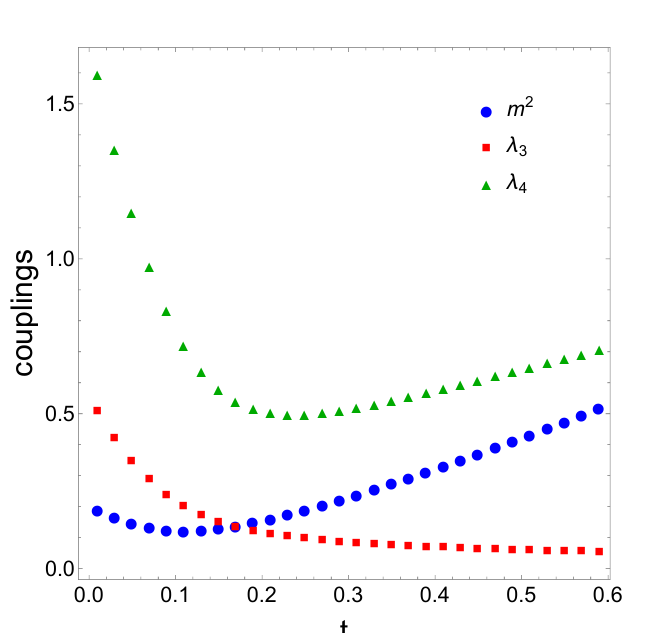}
       \includegraphics[width=0.7\linewidth]{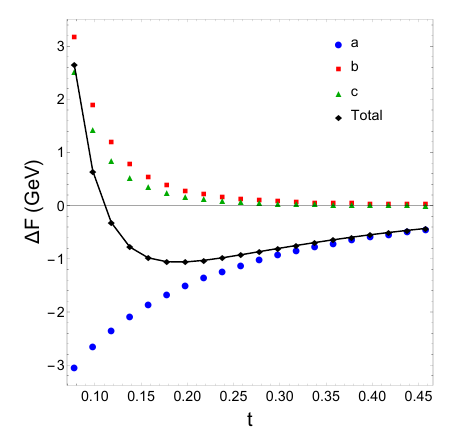}
    \caption{(upper) The quadratic, cubic and quartic couplings as a
    function of reduced temperature. (Lower) Effective free energy of a 4-nucleon cluster. Points marked $a,b,c$ are contributions from diagrams in Fig.1, the contribution of the diagram $d$ turned out to be negligible. }
    \label{fig_couplings.pdf}
\end{figure}

In \cite{DeMartini:2020anq} we used lattice results for critical fluctuations near the second-order phases transitions, as well as analytic results
for effective potential worked out by Heidelberg group via
Wetterich's functional equation. Needless to say, we found good agreement between those.

The derived temperature dependence of 
corresponding couplings is shown
in Fig.\ref{fig_couplings.pdf}, versus the 
reduced temperature 
$$ t={T-T_c \over T_c }$$
Note that
 coefficient of $\phi^2/2$ (squared effective mass $m^2$) is not going to zero at  $t=0$, the CP, although the correlation length is infinite there. 
 Triple and quartic coupling, on the other hand, rapidly grow towards the CP $t\rightarrow 0$.  As a result,  effective free energy $\Delta F$ of a 4-nucleon cluster 
 (lower plot Fig.\ref{fig_couplings.pdf}) changes sign and $\Delta F$
 gets strongly repulsive in the CP vicinity. 

The resulting predictions of that paper were:\\
(i) four-nucleon preclustering $\sim exp(-\Delta F/T)$
should be $suppressed$ in the narrow vicinity of the CP, at $|t|< 1/10$ or so.\\
(ii) This should be observable via multiplcity cumulants, e.g. the kurtosis  $C_4/C_2$ ratio\\
(iii) Similar effect should also be seen in tritium production, because 4-N clusters decay into shallow $O(50)$
states of $^4He$ which have large branchings into $t+p$,
see \cite{Shuryak:2018lgd} for details.

\section{Experimental data}
\begin{figure}
    \centering
   \includegraphics[width=0.7\linewidth]{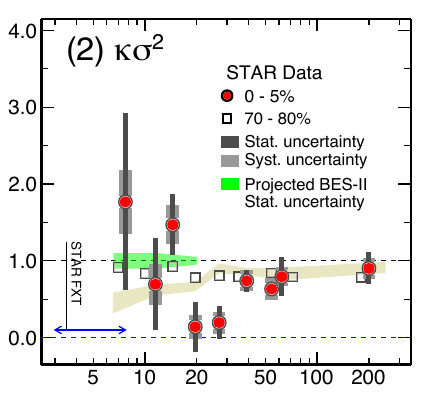}
   \includegraphics[width=0.7\linewidth]      {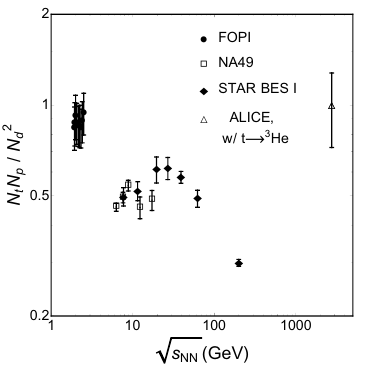}
    \caption{Caption}
    \label{fig_old_data}
\end{figure}

In 2020, when the last of these theory papers \cite{DeMartini:2020anq}  was written, the 
pertinent data were those shown in Fig.\ref{fig_old_data}.
The error bars from BES-I program were too large,
yet curious correlated dips around collision energy $20\, GeV/N$  in $both$ plots were noticed and discussed.

\begin{figure}[h!]
    \centering
 \includegraphics[width=0.9\linewidth]{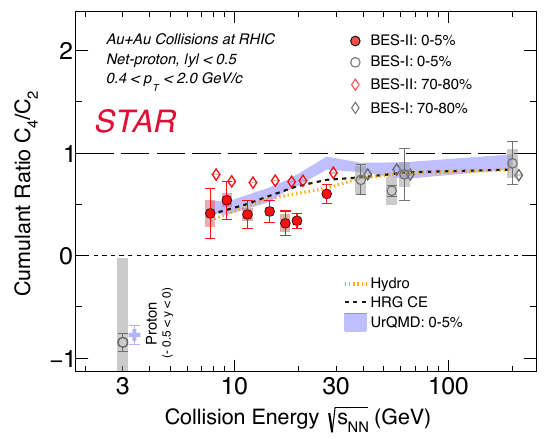}
 \includegraphics[width=0.9\linewidth]{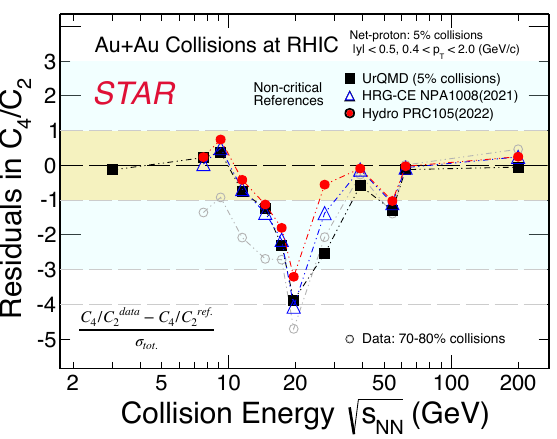}
    \caption{Caption}
    \label{fig_kurtosis}
\end{figure}

At recent Critical Point and Onset of Deconfinement (CPOD 2024, Berkeley)  conference new set of data, from BES-II  run and STAR collaboration
have been presented \cite{BES-II}, see Fig.\ref{fig_kurtosis}.
The accuracy of STAR data are now greatly improved.
Also predictions of conventional models (not possessing the critical fluctuations) were worked out by a number of
approaches. The deviation between the data and these
conventional predictions reveal
 a narrow dip in curtosis,  at energy $\approx 20 \,GeV/N$,
 shown in the lower plot.
 
\section{Conclusions}
(i) The most striking feature is the $sign$ of the 
deviation effect: what is observed is
significant $suppression$ of kurtosis, not an anticipated enhancement in earlier works.
According to  \cite{DeMartini:2020anq}
it is due to $repulsive$ manybody forces from diagrams $(b,c)$
predicted to be dominant close to CP.
\\
(ii) Furthermore, this dip is rather narrow, corresponding to small reduced temperature $|t|<1/10$. (Note that by
$t$ we mean that in effective Ising model. In heavy ion collisions the Ising plot should be rotated, to tangent to the critical line, so it is in fact combination of the temperature and  chemical potential.)\\
(iii)  The dip is located near energy $20 \, GeV/N$, same place
as suspected previously in another dip in (normalized) tritium production ratio.

We end with a general theoretical  argument  from \cite{DeMartini:2020anq} of
 the $sign$ of the effect, which is in fact independent on the details and mandatory. If only attractive binary forces (diagram $a$)
be included, one would have the following paradox: as hypothetically 
the path hits dirctly the CP and thus
the correlation length gets as large as to
reach the size of the fireball, with $N=O(200)$ nucleons
and $N^2/2$ pairs, it would produce so strong mutual attraction of them that fireball would implode, as during a Black Hole formation from a star. Nothing like that is seen, so repulsive multibody forces
 are mandatory in order to prevent that from happening.

The $width$ of the predicted dip, on the other hand, 
depends on numerical values of the couplings. We get those
from universality arguments and solutions to Wetterich
equations for Ising model. We have not performed rotation
of the Ising map to $T-\mu$ plot or fitted any parameters
to kurtosis data yet, but in can soon be done.

Finally,  one may start
discussions of the $shape$ of this dip. For each collision energy
the corresponding adiabatic paths around the CP are different,  so the effective couplings on their paths should in general
deviate, perhaps making the dip  
 somewhat asymmetric. New BES-II data are now good enough 
 on the low energy side, but not yet on the high energy side.
Still this issue should excite theory discussions. \\

As for the preclustering of four nucleons and its contribution
to tritium production, a lot of work is needed to separately
measure three components of the ratio, and see if there is a dip
in tritium component. One can also perhaps hunt for low binding excited states
of $^4He$ via $t+p$ or  $d+d $ resonances.
\bibliography{main}
\end{document}